\documentclass[english,prl,twocolumn,showpacs,superscriptaddress,floatfix,longbibliography]{revtex4-1}
\usepackage[T1]{fontenc}
\usepackage[latin9]{inputenc}
\usepackage{color}
\usepackage{bm}
\usepackage{amsmath}
\usepackage{graphicx}
\usepackage{esint}
\usepackage{babel}
\usepackage{caption}
\usepackage{subcaption}
\begin{document}

\title{Manifestation of extrinsic spin Hall effect in superconducting structures:
Non-dissipative magnetoelectric effects }

\author{F. Sebastian Bergeret}

\affiliation{Centro de Física de Materiales (CFM-MPC), Centro Mixto CSIC-UPV/EHU,
Manuel de Lardizabal 4, E-20018 San Sebastián, Spain}

\affiliation{Donostia International Physics Center (DIPC), Manuel de Lardizabal
5, E-20018 San Sebastián, Spain}

\author{Ilya V. Tokatly}

\affiliation{Nano-Bio Spectroscopy group, Dpto. Física de Materiales, Universidad
del País Vasco, Av. Tolosa 72, E-20018 San Sebastián, Spain }

\affiliation{IKERBASQUE, Basque Foundation for Science, E-48011 Bilbao, Spain}
\begin{abstract}
We present a comprehensive quasiclassical approach for studying transport
properties of superconducting diffusive hybrid structures in the presence
of extrinsic spin-orbit coupling. We derive a generalized Usadel equation
and boundary conditions that in the normal state reduce to the drift-diffusion
theory governing the spin-Hall effect in inversion symmetric materials.
These equations predict the non-dissipative spin-galvanic effect,
that is the generation of supercurrents by a spin-splitting field,
and its inverse -- the creation of magnetic moment by a supercurrent.
These effects can be seen as counterparts of the spin-Hall, anomalous
Hall and their inverse effects in the superconducting state. Our theory
opens numerous possibilities for using superconducting structures
in magnetoelectronics. 
\end{abstract}
\maketitle
The spin-orbit coupling (SOC) in normal systems is at the basis of
striking magnetoelectric effects, such as the spin (SHE)\citep{sinova2015spin}
and anomalous (AHE)\citep{nagaosa2010anomalous} Hall effects widely
studied in normal systems \citep{Zutic2004}. What are the counterpart
of these effects in the superconducting state is, in several aspects,
still an open question. 

According to its origin the SOC can be classified as intrinsic or
extrinsic. Intrinsic SOC generates from the crystal potential associated
with the electronic band structure, and in superconducting structures,
in analogy with the normal state, might lead to non-dissipative magnetoelectric
and spin-galvanic effects as shown in theoretical studies  \citep{Edelstein1995,Bergeret2014a,Konschelle2015,Malshukov2008,Malshukov2010}.
In contrast, extrinsic SOC originates from a random potential due
to impurities. Its influence on the thermodynamics of bulk superconductors
was studied long ago by Abrikosov and Gorkov (AG) \citep{abrikosov1962spin},
who explained non-vanishing magnetic susceptibility of superconductors
at zero temperature. The AG model has been used later to describe
the physics of superconductor-ferromagnet (S-F) structures with SOC.
Within this model, the SOC acts only as a relaxation term for the
spin in the normal and for triplet correlations in the superconducting
state. The suppression of triplet correlations in S-F-S junctions
is associated with the suppression of oscillatory behavior of the
critical Josephson current \citep{demler_arnold_beasley.1997,Bergeret2005}. 

It is well established in the theory of normal systems that SOC not
only leads to spin relaxation, but also to the coupling between spin
and charge currents, responsible for extrinsic SHE and AHE. One expects
that this coupling translates to a singlet-triplet coupling in the
superconducting state, by analogy to the case of non-centrosymmetric
superconductors with intrinsic SOC\citep{Bergeret2014}. However,
for superconductors with extrinsic SOC this coupling has never been
considered, and there is no theoretical framework for its description. 

In this letter we address this issue and derive from a microscopic
model a diffusion equation for superconducting structures with extrinsic
SOC. This equation, Eq.\eqref{eq:Usadel}, generalizes the well known
Usadel equation and contains not only the usual relaxation term due
to the SOC, but also a coupling between spin and charge degrees of
freedom that lead to the SHE and AHE in the normal case. By using
the derived equations we demonstrate that the charge-spin coupling
indeed translates in the superconducting state into singlet-triplet
coupling. Moreover, our equations also show that the lack of a macroscopic
inversion symmetry due, for example, to the presence of hybrid interfaces,
leads to magnetoelectric effects. An example of these is a magnetic
moment induced by a supercurrent. Inversely, SOC leads to the creation
of a supercurrent when the system is polarized via the exchange field
$h$ of a ferromagnet. In the latter case the magnitude of the induced
supercurrent is, as the anomalous Hall voltage, proportional to $h\theta$,
where $\theta$ is the SH-angle. 

\textit{\textcolor{black}{Basic equations for diffusive superconductors
with extrinsic SOC.- }}\textcolor{black}{We first explain how to derive
the generalized Usadel equation and boundary conditions that allow
for an accurate description of superconducting diffusive structures
with extrinsic SOC\cite{supp}. Following the standard derivation of the quasiclassical
equations (see e.g. \citep{Langenberg1986}) the starting point is
the kinetic equation for the Wigner transformed Keldysh $8\times8$
matrix Green's function $\check{G}({\bf p},{\bf r};t,t')$,
\begin{equation}
\tau_{3}\partial_{t}\check{G}+\partial_{t'}\check{G}\tau_{3}+\frac{p_{k}}{m}\partial_{k}\check{G}+i\left[\mathbf{h}\bm{\sigma}\tau_{3}+\check{\Delta},\check{G}\right]={\cal I}\label{eq:kin_equation}
\end{equation}
where ${\bf h}$ is the spin-splitting field, $\check{\Delta}$ is
the anomalous self-energy (SE) describing superconducting correlations,
$\tau_{j}$ and $\sigma^{a}$ are Pauli matrices spanning the Nambu
and spin spaces, respectively. The collision integral ${\cal I}$
in Eq.\eqref{eq:kin_equation} describes scattering at impurities,
\begin{equation}
{\cal I}=-i\left[\check{\Sigma},\check{G}\right]+\frac{1}{2}\left\{ \nabla_{{\bf r}}\check{\Sigma},\nabla_{{\bf p}}\check{G}\right\} -\frac{1}{2}\left\{ \nabla_{{\bf p}}\check{\Sigma},\nabla_{{\bf r}}\check{G}\right\} \label{eq:I-collision}
\end{equation}
where we performed the standard gradient expansion. We describe impurities
by an operator $\hat{W}({\bf r})=V({\bf r})+\hat{V}_{so}(\mathbf{r})$,
with $V({\bf r})$ being a random scalar potential, $\hat{V}_{so}=-i\lambda^{2}\left(\nabla V(\mathbf{r})\times\nabla\right)\bm{\sigma}$
the SOC term, and $\lambda$ the effective Compton wavelength. Within
the Born approximation the SE $\check{\Sigma}({\bf p},{\bf r})$ in
Eq.\eqref{eq:I-collision} is the Wigner transform of $\check{\Sigma}({\bf r_{1},{\bf r_{2})}}=\langle\hat{W}({\bf r_{1})}\check{G}({\bf r_{1},{\bf r_{2}})}\hat{W}({\bf r_{2}})\rangle,$
where the angular brackets denote averaging over impurities configuration.
In $\check{\Sigma}$ we identify two types of terms: (i) those quadratic
in the potentials, $\langle VGV\rangle$ and $\langle\hat{V}_{so}G\hat{V}_{so}\rangle$,
which lead to the relaxation of momentum and spin, respectively, and
(ii) the mixed terms $\langle VG\hat{V}_{so}\rangle$ that account
for the charge-spin coupling. The last terms are traditionally disregarded
in the quasiclassical kinetic theory of superconductors \citep{Alexander1985,demler_arnold_beasley.1997,bergeret_volkov_efetov.2007}.
The importance of mixed terms has been recognized in the context of
spin transport in normal conductors s \citep{RaiSch2010,Shen2014}
where they are responsible for the extrinsic SHE and the spin current
``swapping''. Our goal is to incorporate these magnetoelectric effects
into the quasiclassical theory of diffusive superconductors, which
requires reconsideration of the standard derivation procedure of the
quasiclassical equations. }

\textcolor{black}{To consistently catch the charge-spin coupling one
needs to include gradient terms in the collision integral Eq.\eqref{eq:I-collision}.
This brings momentum derivatives of the GF which do not allow for
a straightforward integration over the particle energy $\xi_{p}$
to derive the Eilenberger equation for the quasiclassical GF }$\check{g}(\mathbf{n})=\frac{i}{\pi}\int d\xi_{p}\check{G}$
that depends on the direction $\mathbf{n}=\mathbf{p}/p_{F}$ of the
Fermi momentum. In order to overcome this difficulty we first obtain
from Eq.\eqref{eq:kin_equation} equations for the zeroth $\sum_{\mathbf{p}}\check{G}$
and first $\sum_{\mathbf{p}}\mathbf{p}\check{G}$ moments of the GF.
At this level one can introduce the quasiclasical GF and consider
directly the diffusive limit in which $\check{g}(\mathbf{n})$ is
approximated as $\check{g}(\mathbf{n})\to\check{g}+n_{k}\check{g}_{k}$,
where $\check{g}$ is the isotropic part and $\check{g}_{k}\ll\check{g}$
is the leading anisotropic correction. The anisotropic part $\check{g}_{k}$
determines the ``matrix current''
\begin{equation}
\check{J}_{k}=v_{F}g_{k}-\frac{\lambda^{2}p_{F}}{4\tau}\epsilon_{kja}\left\{ \sigma^{a},\left[g,g_{k}\right]\right\} ,\label{eq:matrix_J-def}
\end{equation}
where the second term is the ``anomalous velocity'' contribution
due to SOC, and $\tau$ is the momentum scattering time. The physical
charge and spin currents are obtained from the Keldysh component of
the matrix current, $j_{k}=e\pi N_{0}{\rm Tr}\tau_{3}\check{J_{k}^{K}}(t,t)/4$
and $j_{k}^{a}=\pi N_{0}{\rm Tr}\sigma^{a}\check{J}_{k}^{K}(t,t)/4$,
respectively. In the diffusive limit one can solve the equation for
the 1st moment and one finds the anisotropic component $\check{g}_{k}$
that allows to express the matrix current in terms of the isotropic
part $\check{g}$ of GF 
\begin{equation}
\check{J}_{k}=-D\left(\check{g}\partial_{k}\check{g}-\frac{\theta}{2}\epsilon_{kja}\left\{ \sigma^{a},\partial_{j}\check{g}\right\} +i\frac{\kappa}{2}\epsilon_{kja}\left[\sigma^{a},\check{g}\partial_{j}\check{g}\right]\right)\;.\label{eq:matrix_J}
\end{equation}
Here $D$ is the diffusion coefficient. In addition to the usual diffusion
current, Eq. \eqref{eq:matrix_J} contains the expected SH-angle $\theta=2\lambda^{2}p_{F}/v_{F}\tau$
and the swapping term $\kappa=2\lambda p_{F}^{2}/3$ first described
in Ref.\citep{lifshits2009swapping}. From the equation for 0th moment
of the full GF we find that the isotropic component of the GF subjected
to the normalization condition $\check{g}^{2}=1$, satisfies the generalized
Usadel equation,

\begin{eqnarray}
\tau_{3}\partial_{t}\check{g}+\partial_{t'}\check{g}\tau_{3}+\partial_{k}\check{J}_{k}+i\left[\mathbf{h}\bm{\sigma}\tau_{3}+\check{\Delta},\check{g}\right]\label{eq:Usadel}\\
=-\frac{1}{8\tau_{so}}\left[\sigma^{a}\check{g}\sigma^{a},\check{g}\right]+\frac{1}{4}D\theta\epsilon_{kja}\left[\sigma^{a},\check{g}\partial_{k}\check{g}\partial_{j}\check{g}\right]\nonumber 
\end{eqnarray}
where $1/\tau_{so}=8\lambda^{4}p_{F}^{4}/9\tau$. Finally, the Kupriyanov-Lukichev
boundary conditions \citep{Kupriyanov1988} at the interface between
a conventional BCS superconductor and a metal with extrinsic SOC can
be easily generalized by using the matrix current of Eq.\eqref{eq:matrix_J},
\begin{equation}
\nu_{k}\check{J}_{k}=\frac{D}{2R_{b}\sigma_{0}}\left[\check{g}_{BCS},\check{g}\right]\;,\label{eq:K-L-1}
\end{equation}
where $\bm{\nu}$ is a unit vector normal to the interface, $R_{b}$
is the barrier resistance per area, $\sigma_{0}$ the conductivity
of the normal region and $\check{g}_{BCS}$ is the bulk superconductor
GF.

Equations \eqref{eq:matrix_J}-\eqref{eq:K-L-1} are the main results
of this paper. They describe the proximity effect in materials with
extrinsic SOC. Despite the derivation relies on Born approximation,
where only the side-jump contribution to the SH-angle appears \textcolor{black}{\citep{RaiSch2010,Shen2014}},
the final set of Eqs.\eqref{eq:matrix_J}-\eqref{eq:K-L-1} is expected
to be quite general with $\theta$ and $\kappa$ being the material
parameters accounting for all extrinsic and intrinsic (in cubic materials)
contributions to the charge-spin coupling. In fact, Eq.\eqref{eq:matrix_J}
can be viewed as a symmetry based gradient expansion of the current.

In the normal state the terms proportional to $\kappa$ and $\theta$
vanish from Eq.\eqref{eq:Usadel}. These nonlinear in $\check{g}$
terms do appear only if superconducting correlations are present and
may lead to new interesting unexplored phenomena. 

\textit{Non dissipative magnetoelectric effects.-} We now discuss
physical effects predicted by Eqs.\eqref{eq:Usadel}-\eqref{eq:K-L-1}.
For clarity we assume a weak superconducting proximity effect and
linearize the Usadel equation. Moreover, we focus here on non-dissipative
physics, and switch to the equilibrium Matsubara formalism by replacing
in Eq.\eqref{eq:Usadel} $\partial_{t}\to\omega=\pi T(2n+1)$, the
Matsubara frequency. After linearization $\check{g}={\rm sgn}(\omega)\tau_{3}+i\tau_{2}\hat{f}$
the Usadel equation in non-superconducting regions reads 
\begin{equation}
D\nabla^{2}\hat{f}-\left\{ [|\omega|+i{\bf h}\bm{\sigma}\mathbf{{\rm sgn}(\omega)],}\hat{f}\right\} =\frac{3\hat{f}-\sigma^{a}\hat{f}\sigma^{a}}{4\tau_{so}}\;,\label{eq:lin_Usadel}
\end{equation}
where $\hat{f}=f_{s}+{\rm sgn}(\omega)\sigma^{b}f_{t}^{b}$ is the
anomalous GF which describes the induced superconducting condensate
and consists of the singlet $f_{s}$ and odd-frequency triplet $f_{t}^{b}$
components. The linearized boundary condition (\ref{eq:K-L-1}) now
reads
\begin{eqnarray}
\nu_{k}\left(\partial_{k}f_{s}-{\rm \theta}\epsilon_{kja}\partial_{j}f_{t}^{a}\right)=i\gamma f_{BCS}\label{eq:lin_KL-s}\\
\nu_{k}\left(\partial_{k}f_{t}^{a}-{\rm \theta}\epsilon_{kja}\partial_{j}f_{s}-\kappa[\partial_{a}f_{t}^{k}-\delta_{ka}\partial_{j}f_{t}^{j}]\right) & = & 0\label{eq:lin-KL-t}
\end{eqnarray}
where $\gamma=1/R_{b}\sigma_{0}$ and $f_{BCS}=\Delta/\sqrt{\omega^{2}+\Delta^{2}}.$
As we can see from Eqs.\eqref{eq:lin_Usadel}-\eqref{eq:lin-KL-t}
the effect of SOC is twofold. On the one hand, the extrinsic SOC leads
to the known additional relaxation of the condensate (via the Elliot-Yaffet
mechanism), described by the right hand side of Eq.\eqref{eq:lin_Usadel},
if the triplet component is non vanishing. On the other hand, the
SOC induces, out of the singlet, the triplet component at the hybrid
interfaces, even in the absence of the exchange field ${\bf h}$.
The term in Eqs. (\ref{eq:lin_KL-s}), \eqref{eq:lin-KL-t} proportional
to the SH-angle describes the singlet-triplet conversion, which is
the analog to the charge-spin conversion in normal metals. This conversion
can be understood as a consequence of inversion asymmetry at the hybrid
interface. Due to the antisymmetric tensor $\epsilon_{jka}$ in the
SH term the singlet-triplet conversion occurs only in setups with
currents flowing parallel to the interfaces, as for example lateral
Josephson junctions that will be discussed below. 

As a first example\textit{ }we consider a superconducting film with
extrinsic SOC in the absence of the exchange field, ${\bf h}=0$.
The film occupies the region $-d/2<z<d/2$ and is infinite in the
$(x,y)$-plane. The region$z>|d/2|$ is occupied by vacuum and hence 
the boundary condition at $z=\pm d/2$ is obtained by assuming zero current, {\it i.e.} the
 r.h.s of Eq.(\ref{eq:K-L-1}) vanishes.
We assume a small gradient of the superconducting
phase $\nabla\varphi=q\hat{\bm{x}}$ along $x$, so that the singlet
component of the anomalous GF is given by $f_{s}(x)\approx if_{BCS}e^{i\varphi(x)}$.
The triplet component can be easily obtained from Eq.\eqref{eq:lin_Usadel}
and  Eq.\eqref{eq:lin-KL-t} which for the
present geometry read $\partial_{z}f_{t}^{y}\vert_{\pm d/2}=\theta\partial_{x}f_{s}\approx-\theta qf_{BCS}$.
Despite the film is nonmagnetic (${\bf h}=0$), the $y$-component
of the triplet is generated due to a finite SH-angle $\theta$, and
this leads to a finite magnetic moment $m^{y}=\mu_{B}2\pi N_{0}T\sum_{\omega}{\rm Im}\left[f_{s}^{*}f_{t}^{y}\right]$:

\begin{equation}
m^{y}(z)=\mu_{B}\theta T\sum_{\omega}\frac{j_{x}(\omega)}{Dk}\frac{\sinh kz}{\cosh kd/2}\quad,\label{eq:my_Sfilm}
\end{equation}
where $j_{x}(\omega)=q\pi DN_{0}f_{BCS}^{2}$ is the spectral supercurrent,
and $k^{2}=k_{\omega}^{2}+k_{so}^{2}$ with $k_{\omega}^{2}=2|\omega|/D$
and $k_{so}=1/D\tau_{so}$. The induced magnetization Eq.\eqref{eq:my_Sfilm}
is opposite at opposite sides of the film so that the net magnetic
moment is zero, which is a clear consequence of the inversion symmetry.
The supercurrent-induced accumulation of the odd-frequency triplet
component and the spin density at the film edges is the non-dissipative
analog of extrinsic SHE. 

Let us now consider a  normal metal layer (N)  of thickness $d$  
and  finite SH-angle $\theta$, in contact with a bulk superconductor.
The N and S  layers  occupies the region $0<z<d$  and  $z<0$ respectively. 
We assume a supercurrent flowing within
the S layer due to a small phase gradient $\nabla\varphi=q\hat{\bm{x}}$.
Because of the proximity effect the singlet component penetrates N
where it is converted to a triplet component due to the SH term in
the boundary conditions. Both singlet and triplet components can be
easily determined from Eq.(\ref{eq:lin_Usadel}) and the boundary
conditions at the S/N interface Eqs.(\ref{eq:lin_KL-s}), \eqref{eq:lin-KL-t}.
The induced magnetic moment is then given by: 
\begin{eqnarray*}
m^{y}(z) & = & \mu_{B}\gamma^{2}\theta T\sum_{\omega}\frac{j_{x}(\omega)\cosh k_{\omega}(z-d)}{Dk_{\omega}^{2}k\sinh^{2}k_{\omega}d\sinh kd}\\
 & \times & \left[\cosh kz-\cosh k_{\omega}d\cosh k(z-d)\right]
\end{eqnarray*}
Thus the supercurrent flowing in the S layer induces a spin density
over the whole N layer. In contrast to our previous example, now the
net magnetization is nonzero. In other words, the supercurrent generates
a global spin, which is allowed due to the structure inversion asymmetry
of the S/N bilayer. Phenomenologically this can be described as a
non-dissipative Edelstein effect (EE). The important difference with
the usual EE \citep{Edelstein1995,Bergeret2014a,Konschelle2015} is
that it originates solely from the extrinsic SOC and the macroscopic
asymmetry of the structure.

\begin{figure}
\includegraphics[width=1\columnwidth]{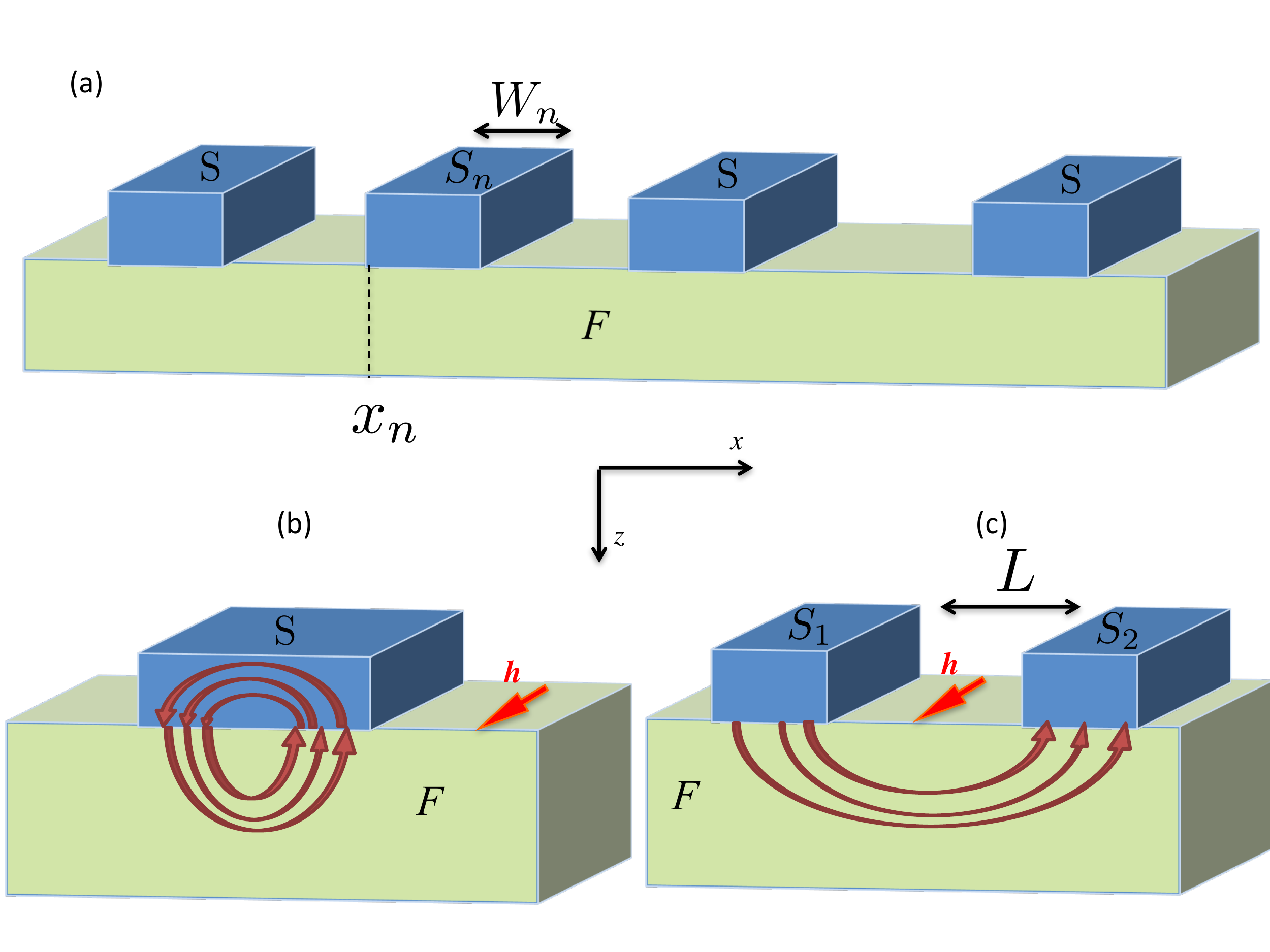}

\caption{Lateral S/F structures\label{fig:Lateral-S/N} and illustration of
the supercurrent flow.}
\end{figure}

Experimentally it might be easier to detect the inverse of this effect.
Namely, the generation of supercurrents by a combination of SOC and
exchange field, which is our third example. We consider a multi-terminal
lateral S/F structure (Fig. \ref{fig:Lateral-S/N}) which resembles lateral structures used in experiments on SFS 
structures\cite{keizer_goennenwein_klapwijk_et_al.2006,Wang2010,Anwar2010,PhysRevX.5.021019}
 The $n$-th
S terminal is infinite in $y$-direction and has a width $W_{n}$,
while F is a ferromagnet with an exchange field ${\bf h}=h\hat{\bm{y}}$
along $y$. The current density flowing through the $n$-th S/N interface
is readily obtained from Eq.\eqref{eq:K-L-1}: 
\begin{equation}
j_{z}^{(n)}(x)=\frac{\pi T}{eR_{bn}}\sum_{\omega}{\rm Im}\left[f_{0}^{(n)}(x)f_{s}^{*}(x,0)\right],\label{eq:lateral_current-1}
\end{equation}
here $f_{0}^{(n)}(x)=if_{BCS}e^{i\varphi_{n}}\left[\Theta(x-x_{n})-\Theta(x-x_{n}-W_{n})\right]$
is the GF of the $n$-th S electrode with the phase $\varphi_{n}$,
and $f_{s}(x)$ is the singlet component induced in N at $z\to0$.
If all phases are identical, e.g. $\varphi_{n}=0$ for all $n$, only
the real part $f_{s}^{{\rm Re}}(x)$ of the singlet GF in N contributes
to the current as in this case ${\rm Im}[f_{0}^{(n)}f_{s}^{*}]=f_{BCS}f_{s}^{{\rm Re}}$.
From Eqs.\eqref{eq:lin_Usadel}-\eqref{eq:lin-KL-t} we find that
only for simultaneously non-vanishing $h$ and $\theta$ the component
$f_{s}^{{\rm Re}}(x)$ can be generated as follows: Due to the proximity
effect a purely imaginary ``primary'' $f_{s}$ is induced in F,
where it is converted, via the exchange coupling term $h$ in Eq.\eqref{eq:lin_Usadel},
into the real triplet $f_{t}^{{\rm Re}}$. Finally, $f_{t}^{{\rm Re}}$
is converted into $f_{s}^{{\rm Re}}$ via the SH term, $\theta$,
in Eq.\eqref{eq:lin_KL-s}. Since the SH singlet-triplet coupling
involves gradients, it is clear that $f_{s}^{{\rm Re}}(x)$ will be
generated near inhomogeneities -- the edges of the S terminals. In
general the function $f_{s}^{{\rm Re}}(x)$ can be written as follows
\begin{equation}
f_{s}^{{\rm Re}}(x)=k_{h}^{2}\theta\sum_{n=1}^{M}\gamma_{n}\left[s(x-x_{n})-s(x-x_{n}-W_{n})\right]\label{eq:lateral_fs1}
\end{equation}
where $M$ is the number of terminals, $k_{h}^{2}=2h/D$, $\gamma_{n}=1/R_{bn}\sigma_{0}$,
and $s(x)$ is a function localized near the origin and describing
the singlet component induced at the left/right edges of each S electrode.
In the limit of thick, formally semi-infinite F layer we find (see
SM for details)
\begin{eqnarray*}
\frac{s(x)}{f_{BCS}} & = & \frac{k_{-}^{2}-k_{\omega}^{2}}{k_{+}(k_{+}^{2}-k_{-}^{2})^{2}}e^{-k_{+}|x|}+\frac{k_{+}^{2}-k_{\omega}^{2}}{k_{-}(k_{+}^{2}-k_{-}^{2})^{2}}e^{-k_{-}|x|}\\
 & - & \frac{2k_{so}^{2}}{\pi^{2}(k_{+}^{2}-k_{-}^{2})^{2}}\int_{-\infty}^{\infty}dx'K_{0}(k_{+}|x-x'|)K_{0}(k_{-}|x'|)
\end{eqnarray*}
where $k_{\pm}^{2}=k_{\omega}^{2}+k_{so}^{2}/2\pm\sqrt{k_{so}^{4}/4-k_{h}^{4}}$,
and $K_{0}(x)$ is the modified Bessel function of the second kind. 

In the one-terminal case ($M=1$) the right hand side in Eq. (\ref{eq:lateral_fs1})
is antisymmetric with respect to the center of the terminal. Therefore
the current $j_{z}^{(1)}(x)$ Eq.\eqref{eq:lateral_current-1}, being
also antisymmetric, averages to zero after the integration over $x$.
In other words, in a one-terminal S/N lateral structure, the combination
of the extrinsic SOC and the exchange field generates circulating
currents as sketched in Fig. \ref{fig:Lateral-S/N}b. 

In the two-terminal case, shown on Fig. \ref{fig:Lateral-S/N}c, the
total current $I_{1}$ flowing through S1-terminal is nonzero due
to $f_{s}^{{\rm Re}}(x)$ induced from S2-terminal:
\begin{eqnarray*}
I_{1} & = & \frac{\pi k_{h}^{2}\theta T}{eR_{b1}R_{b2}\sigma_{0}}\sum_{\omega}f_{BCS}\\
 & \times & \int_{x_{1}}^{x_{1}+W_{1}}dx\left[s(x-x_{2})-s(x-x_{2}-W_{2})\right]
\end{eqnarray*}
Therefore besides currents circulating around each interface, there
is a finite Josephson current induced by mutual effect of extrinsic
SOC and the exchange field (see Fig. \ref{fig:Lateral-S/N}c). This
supercurrent at $\varphi=0$ resembles the anomalous current in a
$\varphi_{0}$-junction studied in the context of intrinsic SOC in
polar crystals \citep{Bergeret2014a,Konschelle2015,Buzdin2008}. 
Here
we show that $\varphi_{0}$-junction can be built out of the most
common inversion symmetric materials provided they show a finite exchange
spin-splitting and a SH-angle. The anomalous current is proportional
to $\theta h$, which in turn is proportional to the anomalous Hall
conductivity $\sigma_{AH}$ in ferromagnets \citep{nagaosa2010anomalous}.
Hence F materials with large $\sigma_{AH}$ are good candidates for
showing an anomalous supercurrent in  lateral SFS structures. If for example one uses
a ferromagnet with strong exchange field such that $k_h^2\gg k_{so}^2$, the amplitude of the 
anomalous current is according to our theory proportional to $\theta$ times the critical current of 
the junction.Thus, for materials with $\theta\sim 5-20\%$  the anomalous phase  current 
can be detected  by using quantum interferometer devices as done for example in Ref.\cite{delft} for nanowires.


\begin{figure*}
    \centering
    \begin{subfigure}[b]{0.5\textwidth}
        \includegraphics[width=\linewidth]{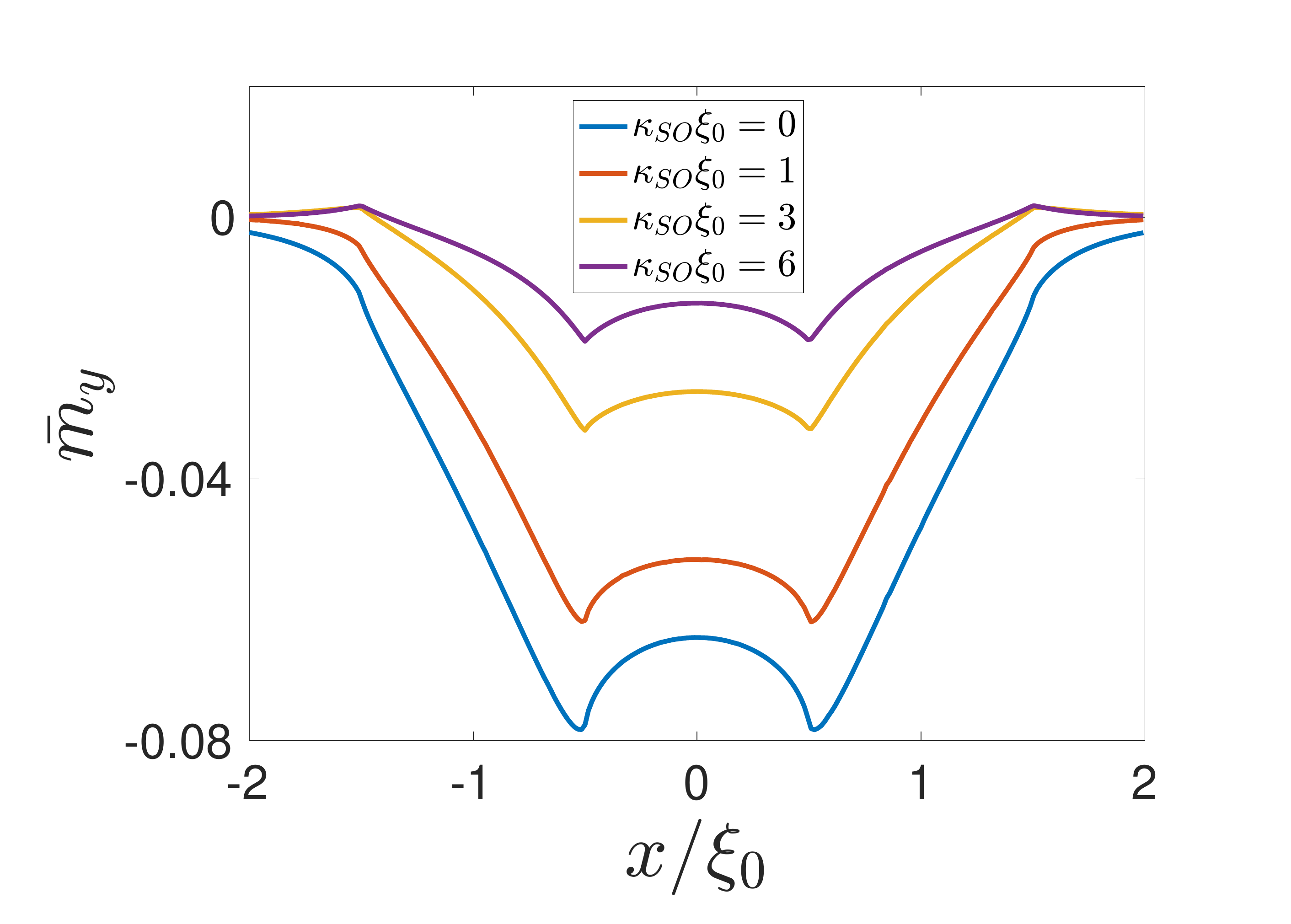} 
    \end{subfigure}
    \hspace{-.5cm}
        \begin{subfigure}[b]{0.5\textwidth}
        \includegraphics[width=\linewidth]{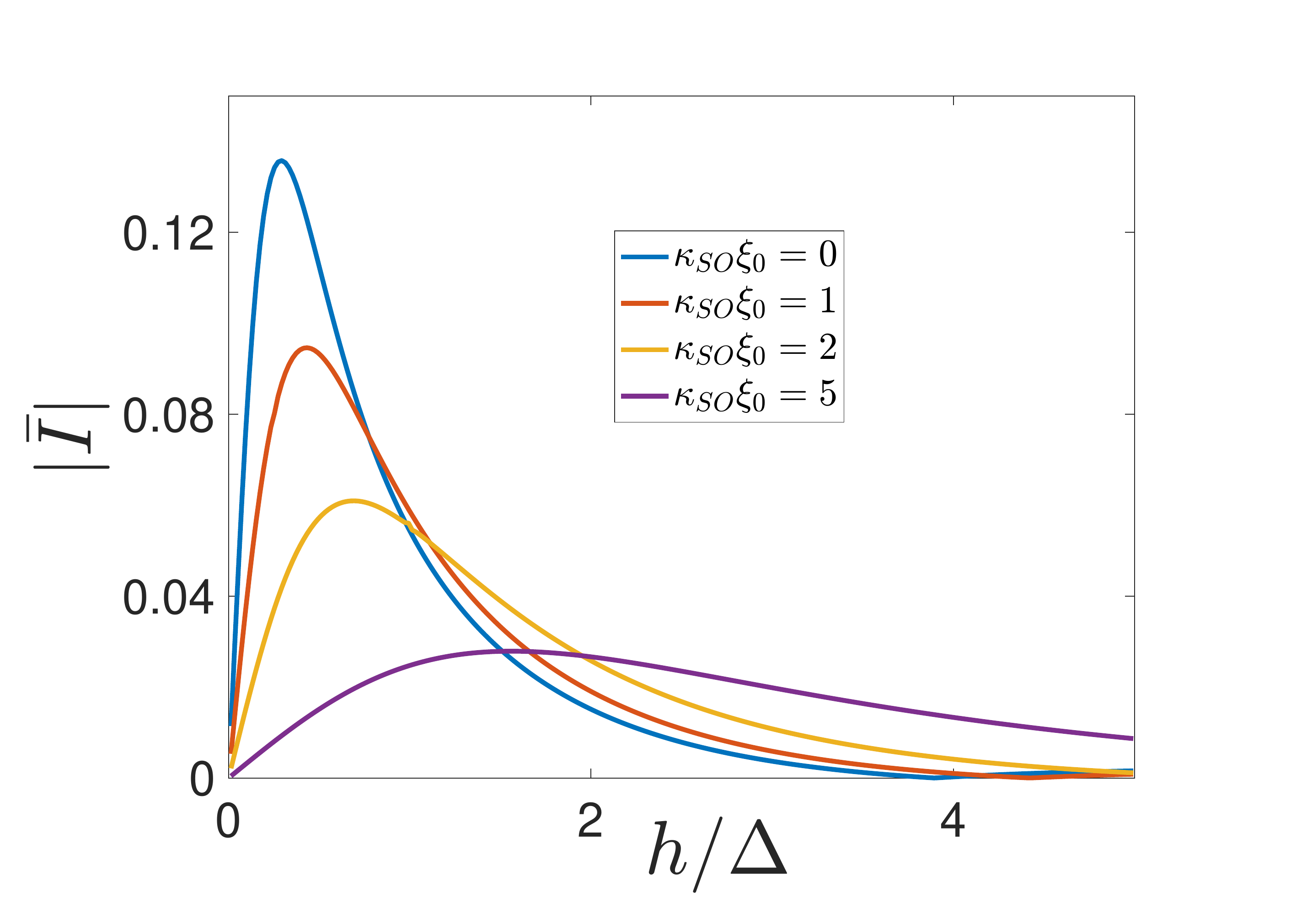} 
    \end{subfigure}
\caption{ Left panel: the x-dependence of the induced magnetic moment $\bar{m}^{y}=m^{y}/(\pi\mu_{B}N_{0}\theta\gamma_{1}\gamma_{2}D)$
at $z=0$ for a symmetric lateral SNS junction with $L=1$, $W=1$
and $\varphi=\frac{\pi}{2}$. Right panel: induced anomalous supercurrent
as a function of the exchange field for $L=1$ and W=5. We have chosen
$T=0.1$. \label{fig:fig3} and defined $\xi_{0}=\sqrt{D/2\Delta}$.}
\end{figure*}

In the right panel of Fig. \ref{fig:fig3} we show the anomalous current
through the S1-terminal as a function of the field $h$.
The current starts from zero at $h=0$, reaches a maximum and finally
decays for large fields because of the usual suppression of superconductivity
 by the $h$ field \citep{buzdin.2005_RMP,Bergeret2005}. 
Inversely, if $h=0$ a finite Josepshon current ($\varphi\neq 0$)
between the two S electrodes induces a finite magnetic moment (see 
SM for details) similar to the situation found in the S and S/N layered
systems. In the left panel of Fig. \ref{fig:fig3} we show the $x$-dependence
of the magnetic moment induced at $z=0$.  

In conclusion, we have presented a new theoretical framework that
describes diffusive superconducting hybrid structures with extrinsic
SOC. We have derived equations that contain hitherto unknown terms
proportional to the SH-angle, responsible in the normal state for
the SHE, and the Lifshits-Dyakonov spin-currents swapping parameter.
Our equations pave the way to explore numerous novel effects
 in the field of superconducting spintronics \cite{Wakamura,Maekawa,Reviews}
 and open up numerous opportunities for the control of charge and spin currents
in the non-dissipative regime. As illustrative examples we demonstrate
the existence of magnetoelectric effects in different superconducting
structures. We show that these effects are proportional to the SH-angle
and hence can be observed by combining materials with known large
$\theta$, like Pt or Co, with superconducting electrodes. 
\begin{acknowledgments}
The work of F.S.B. was supported by Spanish Ministerio de Economía
y Competitividad (MINECO) through the Project No. FIS2014-55987-P
and the Basque Government under UPV/EHU Project No. IT-756-13. I.V.T.
acknowledges support from the Spanish Grant FIS2013-46159-C3-1-P,
and from the \textquotedblleft Grupos Consolidados UPV/EHU del Gobierno
Vasco\textquotedblright{} (Grant No. IT578-13)
\end{acknowledgments}

\bibliographystyle{apsrev4-1}
%


\begin{widetext}
\section{Supplementary Material}

\subsection{Derivation of the Usadel  equation in the presence
of extrinsic spin-orbit coupling}

{In this section we derive 
the generalised  Usadel  equation to account for magnetoelectric effects [Eq. (5) in the main text]. We consider
a diffusive conventional superconductor described by the Hamiltonian
\begin{equation}
H=H_{BCS}-{\bf h}{\bm \sigma}+\hat{W}(\mathbf{r})\;,\label{eq:Hamiltonian}
\end{equation}
where $H_{BCS}$ is the usual mean field BCS Hamiltonian, ${\bf h}$
is the exchange field, ${\bm \sigma}=(\sigma_{1},\sigma_{2},\sigma_{3})$
the Pauli matrices, and  $\hat W({\bf r})$ is a random impurity potential  
\begin{equation}
\hat{W}({\bf r})=V({\bf r})+\hat{V}_{SOC}(\mathbf{r})\; .\label{eq: potW}
\end{equation}
It  consists of the usual (scalar) elastic scattering $V({\bf r})$ and the
spin-orbit part
\begin{equation}
\hat{V}_{SO}=\lambda^{2}\left({\bf \nabla} V(\mathbf{r})\times\hat{\mathbf{p}}\right){\bm{\sigma}}\; , \label{eq:VSOC}
\end{equation}
where the coupling constant is proportional to the effective Compton wavelength
$\lambda$ squared and $\hat{{\bf p}}=-i\nabla_{{\bf r}}$ the momentum
operator. In order to derive the quantum diffusion equation we introduce
the Keldysh matrix Green functions (GF) which is the  $8\times8$ matrix
\[
\check{G}=\left(\begin{array}{cc}
G^{R} & G^{K}\\
0 & G^{A}
\end{array}\right)\;,
\]
consisting of   the retarded, advanced and Keldysh 4$\times$4
matrices  ($G^{R,A,K}$) in the Nambu-spin space. $\check{G}$ obeys the equation\cite{Langenberg1986}
\begin{equation}
\left[\tau_{3}i\partial_{t}+\frac{1}{2m}\partial_{r}^{2}+\mu+{\bf h}{\bm \sigma}+\check{\Delta}-\check{\Sigma}\right]\check{G}({\bf r},t;{\bf r'},t')=\delta({\bf r-r'})\delta(t-t')\;,\label{eq:Gorkov}
\end{equation}
where $\mu$ is the chemical potential, $\check{\Delta}$ the superconducting
order parameter and $\check{\Sigma}$ is the self-energy due to the
impurity scattering, Eq. (\ref{eq: potW}). We treat the latter within
the self-consistency Born approximation, {\it i.e.}
\[
\check{\Sigma}({\bf r_{1}},{\bf r_{2})}=\left\langle \hat{W}({\bf r_{1})\check{G}({\bf r_{1},{\bf r_{2})\hat{W}({\bf r}_{2})}}}\right\rangle =\check{\Sigma_{0}}+\check{\Sigma}_{1}+\check{\Sigma_{2}\;,}
\]
where $\langle...\rangle$ denotes average over the impurity configuration.
The three terms on the r.h.s correspond to the following  contributions:  

The usual elastic scattering term 
\begin{equation}
\Sigma_{0}=\left\langle V({\bf r_{1}})\check{G}({\bf r_{1}},{\bf r_{2}})V({\bf r_{2}})\right\rangle \;,\label{eq:Sigma0}
\end{equation}
the spin relaxation term which, quadratic in the SOC potential 
\begin{equation}
\Sigma_{2}=\left\langle \hat{V}_{SO}({\bf r_{1}})\check{G}({\bf r_{1}},{\bf r_{2}})\hat{V}_{SO}({\bf r_{2}})\right\rangle \label{eq:Sigma2}
\end{equation}
 and the ``mixed'' term 
\begin{equation}
\Sigma_{1}=\left\langle V({\bf r_{1}})\check{G}({\bf r_{1}},{\bf r_{2}})\hat{V}_{SO}({\bf r_{2}})\right\rangle +\left\langle \hat{V}_{SO}({\bf r_{1}})\check{G}({\bf r_{1}},{\bf r_{2}})V({\bf r_{2}})\right\rangle \label{eq:Sigma1}
\end{equation}
which is responsible for the coupling between charge and spin degrees
of freedom and leads to the SHE and AHE.
 As usual, we assume  for the random potential 
\begin{equation}
\langle V({\bf r_{1}})V({\bf r_{2}})\rangle=\frac{1}{2\pi N_{F}\tau}\delta({\bf r_{1}}-{\bf r_{2}})\;,\label{eq:imp_corr}
\end{equation}
where $\tau$ is the momentum relaxation time.}

\textcolor{black}{We follow the usual steps in order to obtain the
quantum kinetic equation from Eq. (\ref{eq:Gorkov})\cite{Langenberg1986}: (1) One subtracts
from Eq. (\ref{eq:Gorkov}) its conjugate, (2) performs the Wigner
transform and then (3) the gradient expansion. After
these steps are carried out one obtains the  kinetic-like equation [Eq.
(1) in the main text]: 
\begin{equation}
\frac{p_{k}}{m}\partial_{k}G+i\tau_{3}\partial_{t}G-i\partial_{t'}G\tau_{3}+i\left[\mathbf{h}{\bm \sigma}\tau_{3},G\right]={\cal I}\;,\label{eq:kin_eq}
\end{equation}
where  ${\cal I}=-i\left[\check{\Sigma},\check{G}\right]={\cal I}_{0}+{\cal I}_{1}+{\cal I}_{2}$ is the collision term. It  consists of 
 three contributions 
corresponding to the three self-energy terms (\ref{eq:Sigma0}-\ref{eq:Sigma1}).
${\cal I}_{0}$ and ${\cal I}_{2}$ can be treated in  the
lowest order of the gradient expansion. In contrast and  in order to catch consistently
the charge-spin coupling we need to include linear terms of ${\cal I}_{1}$  in the gradient
expansion:}
\textcolor{black}{
\begin{equation}
{\cal I}_{1}({\bf r},{\bf p})=-i\left[\check{\Sigma}_{1{\bf p}},\check{G}_{{\bf p}}\right]+\frac{1}{2}\left\{ \partial_{{\bf r}}\check{\Sigma}_{1{\bf p}},\partial_{{\bf p}}\check{G}_{{\bf p}}\right\} -\frac{1}{2}\left\{ \partial_{{\bf p}}\check{\Sigma}_{1{\bf p}},\partial_{{\bf r}}\check{G}_{{\bf p}}\right\}\dots \;.\label{eq:col_int}
\end{equation}
}

The derivation of the quasiclassical expressions for  ${\cal I}_{0,2}$
 follows the standard steps, and hence those terms will be added straightforwardly in the end equation. Here we focus on the term ${\cal I}_{1}$ and how to include it in the quasiclassical formalism. 
 
We start by writing explicitly the self-energy Eq. (\ref{eq:Sigma1}):
 \begin{equation}
\Sigma_{1}=-i\langle  A_{j}^{a}({\bf r_{1}})V({\bf r_{2}})\rangle \partial_{{\bf r}_{1}^{j}}\sigma^{a}G({\bf r_{1}},{\bf r_{2}})+i\langle  A_{j}^{a}({\bf r_{2}})V({\bf r_{1}})\rangle \partial_{{\bf r}_{2}^{j}}G({\bf r_{1}},{\bf r_{2}})\sigma^{a}\;,
\end{equation}
where    
\[
A_{j}^{a}({\bf r_{1}})=\lambda^{2}\epsilon_{kja}\partial_{{r_{1}^{k}}}V({\bf r_{1}})\; ,
\]
$\epsilon_{ijk}$ is the Levi-Civita tensor, and sum over repeated indices is implied. 

 By using Eq. (\ref{eq:imp_corr}) one obtains:
\[
\Sigma_{1}=-i\frac{\lambda^{2}}{2\pi N_{F}\tau}\epsilon_{kja}\partial_{{\bf r}_{1}^{k}}\delta({\bf r_{1}-r_{2}})\sigma^{a}\partial_{{\bf r}_{1}^{j}}G({\bf r_{1},r_{2}})+i\frac{\lambda^{2}\epsilon_{kja}}{2\pi N_{F}\tau}\partial_{{\bf r}_{2}^{k}}\delta({\bf r_{1}-r_{2}})\partial_{{\bf r}_{2}^{j}}G({\bf r_{1},r_{2}})\sigma^{a}\;.
\]
Now we  Wigner-transform this expression. This implies  to go over  the  relative ${\bm \rho}{\bf ={\bf r_{1}-r_{2}}}$
and center of mass coordinates ${\bf r}={\bf (r_{1}}+{\bf r_{2}})/2$ and to  Fourier- transform with respect to ${\bm \rho}$:
\begin{equation}
\Sigma_{1}({\bf r},{\bm p})=-i\frac{\lambda^{2}\epsilon_{kja}}{2\pi N_{F}\tau}\sum_{\bm{p'}}\int {\bm {d \rho}} e^{-i {\bm\rho}({\bm p}-{\bm {p'}})}\partial_{\rho^k}\delta({\bm \rho})\left(ip'_{j}\left[\sigma^{a},G(\bm{p'})\right]+\frac{1}{2}\partial_{r^{j}}\left\{ \sigma^{a},G(\bm{p'})\right\} \right)\; .
\end{equation}
By noticing that $\sum_{\bm p} \epsilon_{ija}p_ip_jG^a(\bm{p})=0$ and that the Green's functions are peaked at the Fermi level we can express $\Sigma_1$  in terms of the quasiclassical GFs $g=(i/\pi)\int d\xi G$ as:
\begin{equation}
\Sigma_{1}=\frac{\lambda^{2}\epsilon_{kja}}{2\tau}\left[\sigma^{a},p_{k}\langle p_{j}g\rangle\right]-i\frac{\lambda_{c}^{2}\epsilon_{kja}}{2\tau}\partial_{r^{j}}\left\{ \sigma^{a},p_{k}\langle g\rangle-\langle p_{k}g\rangle\right\} \;.\label{eq:Smix}
\end{equation}
In this last expression the brackets denote average over the momentum
direction. It is important to note that the second term contains
a gradient  and hence it is, in principle, of smaller order than the
first one in the gradient expansion.  As noticed before, the description of  the spin-charge coupling 
compels  to keep these higher order terms. 

We substitute now  Eq. (\ref{eq:Smix})  into the expression for the
collision term Eq. (\ref{eq:col_int}) and keep  terms up to
linear order in the gradients:
\begin{equation}
{\cal I}_{1}\approx-i\left[\Sigma_{1}^{(0)},G_{\bm p}\right]-i\left[\Sigma_{1}^{(1)},G_{\bm p}\right]+\frac{1}{2}\left\{ \partial_{{\bf r}}\Sigma_{1{\bf p}}^{(0)},\partial_{{\bf p}}G_{{\bf p}}\right\} -\frac{1}{2}\left\{ \partial_{{\bf p}}\check{\Sigma}_{1{\bf p}}^{(0)},\partial_{{\bf r}}\check{G}_{{\bf p}}\right\} \;,\label{eq:I1_final}
\end{equation}
 where $\Sigma_{1\bm p}^{(0,1)}$ are  the first and second term in Eq. (\ref{eq:Smix})
respectively.   {{We emphasise once again that in order to get the next-leading order correction correctly it is crucial to keep {\it all} terms in the expansion Eq. (\ref{eq:col_int}).  
}}

The collision term described by  Eq. (\ref{eq:I1_final}) does not allow for a straightforward
integration over the quasiparticle energy and hence one cannot derive 
a closed  differential equation  (Eilenberger equation) for the quasiclassical $\check{g}$. 
In order to overcome this difficulty  we consider the diffusive limit and derive equations for the zeroth
$\sum_{{\bf p}}\check{G}$ and first $\sum_{{\bf p}}{\bf p}\check{G}$
moments of $\check{G}$.

From Eq. (\ref{eq:I1_final}) we obtain :
\begin{equation}
\frac{i}{\pi N_{F}}\sum_{\bm p}{\cal I}_{1}=\epsilon_{kja}\frac{\lambda^{2}}{4\tau}\left(i\left\{ \sigma^{a},\left[\langle  p_{k}g\rangle ,\langle  p_{j}g\rangle \right]\right\} +\partial_{k}\left\{ \sigma^{a},\left[\langle  g\rangle ,\langle  p_{j}g\rangle \right]\right\} -\left[\sigma^{a},\left\{ \langle  g\rangle ,\partial_{k}\langle  p_{j}g\rangle \right\} \right]\right)\;,\label{eq:Iaverag}
\end{equation}
for the zeroth moment of ${\cal I}_{1}$ and  
\begin{equation}
\frac{i}{\pi N_{F}}\sum_{\bm p}p_{k}{\cal I}_{1}=-\epsilon_{kja}\frac{\lambda^{2}p_{F}^{2}}{2\tau}\frac{1}{3}\left(i\left[\left[\sigma^{a},\langle p_{j}g\rangle \right],\langle  g\rangle \right]+\frac{1}{2}\left[\left\{ \sigma^{a},\partial_{j}\langle  g\rangle \right\} ,\langle  g\rangle \right]\right)\label{eq:p_averageImix}
\end{equation}
 for the first moment.

In the diffusive limit one assumes that  
$\tau E\ll1$ and $\lambda^{2}p_{F}^{2}\ll1$(  $E$ is any energy
involved in the kinetic equation) and expands $g$ in spherical
harmonics: $g\approx g_{0}+n_{k}g_{k}$ such that $\langle  g\rangle =g_{0}$, 
$\langle  p_{k}g\rangle =p_{F}g_{k}$, and  $g_{0}\gg g_{k}$,  . In  this limit  one can simplify
expressions (\ref{eq:Iaverag}-\ref{eq:p_averageImix}) and get: 
\begin{equation}
\frac{i}{\pi N_{F}}\sum_{\bm p}{\cal I}_{1}\approx\epsilon_{kja}\frac{\lambda^{2}p_{F}}{4\tau}\left(\partial_{k}\left\{ \sigma^{a},\left[g_{0},g_{j}\right]\right\} -\left[\sigma^{a},\left\{ g_{0},\partial_{k}g_{j}\right\} \right]\right)\label{eq:Imix_iso}
\end{equation}
and\texttt{
\begin{equation}
\frac{i}{\pi N_{F}}\sum_{\bm p}p_{k}{\cal I}_{1}\approx-\epsilon_{kja}\frac{\lambda^{2}p_{F}^{2}}{4\tau}\frac{1}{3}\left(i\left[\left[\sigma^{a},g_{j}\right],g_{0}\right]+\frac{1}{2}\left[\left\{ \sigma^{a},\partial_{j}g_{0}\right\} ,g_{0}\right]\right)\; .\label{eq:Imix_aniso}
\end{equation}
}

By switching to  the Matsubara representation in Eq. (\ref{eq:Gorkov})
we obtain for the zero and first moments 
\begin{equation}
\partial_{k}\left(v_{F}g_{k}-\epsilon_{kja}\frac{\lambda^{2}p_{F}}{4\tau}\left\{ \sigma^{a},\left[g_{0},g_{j}\right]\right\} \right)+\left[(\omega-i{\bf h}{\bm \sigma)\tau_{3},}g_{0}\right]=-\frac{1}{8\tau_{SO}}\left[\sigma^{a}g_{0}\sigma^{a},g_{0}\right]-\epsilon_{kja}\frac{\lambda^{2}p_{F}}{4\tau}\left[\sigma^{a},\left\{ g_{0},\partial_{k}g_{j}\right\} \right]\label{eq:kin_iso-2}
\end{equation}
and 
\begin{equation}
\frac{\tau v_{F}}{3}\partial_{k}g_{0}+\epsilon_{kja}\frac{\lambda^{2}p_{F}}{4}\frac{1}{3}\left[\left\{ \sigma^{a},\partial_{j}g_{0}\right\} ,g_{0}\right]=-\frac{1}{2}\left[g_{0},g_{k}\right]+\epsilon_{kja}\frac{\lambda^{2}p_{F}^{2}}{2}\frac{i}{3}\left[\left[\sigma^{a},g_{j}\right],g_{0}\right]\,,\label{eq:kin_aniso-1}
\end{equation}
where \textcolor{black}{$1/\tau_{SO}=8\lambda^{4}p_{F}^{4}/9\tau$,
and} in the second equation we only took leading order terms in the
diffusive expansion. 
At this stage and before writing the Usadel equation, it is worth to make  two  remarks:  (i) The
first term in Eq. (\ref{eq:kin_iso-2}) is the divergence of the matrix
current 
\begin{equation}
\check{J}_{k}=v_{F}g_{k}-\epsilon_{kja}\frac{\lambda^{2}p_{F}}{4\tau}\left\{ \sigma^{a},\left[g_{0},g_{j}\right]\right\} \;.\label{eq:current0}
\end{equation}
The last term of this expression stems form the SOC and described the coupling between the charge and spin currents. 
(ii) The structure of Eq. (\ref{eq:kin_aniso-1}), $\partial_{k}g_{0}+[A,g_{0}]=0$,
ensures the validity of the normalization condition
\begin{equation}
g_{0}^{2}=1\;.\label{eq:normalization}
\end{equation}

The final  step is to get an expression for the anisotropic component
$g_{k}$ in terms of the isotropic one $g_{0}$ from Eq. (\ref{eq:kin_aniso-1}).
In leading order with respect to the parameters $\lambda^{2}p_{F}^{2}$
and $\lambda^{2}p_{F}/L$, where $L$ is the characteristic length
over which $g_{0}$ varies, the anisotropic component reads
\begin{equation}
g_{k}=-\tau v_{F}g_{0}\partial_{k}g_{0}+\epsilon_{kja}\frac{\lambda^{2}p_{F}}{2}\left\{ \sigma^{a},\partial_{j}g_{0}\right\} -\epsilon_{kja}\lambda^{2}p_{F}^{2}\frac{i}{3}\left[\sigma^{a},\tau v_{F}g_{0}\partial_{j}g_{0}\right]\;.\label{eq:gani}
\end{equation}
This can be checked by substituting Eq. (\ref{eq:gani}) into
Eq. (\ref{eq:kin_aniso-1}), using the normalization condition (\ref{eq:normalization}),  and by  keeping only leading order terms.
If we now substitute this expression for $g_{k}$ into Eq. (\ref{eq:current0})
we obtain the expression of the matrix current in terms of $g_0$:
\begin{equation}
J_{k}=-D\left(g_{0}\partial_{k}g_{0}-\frac{\theta}{2}\epsilon_{kja}\left\{ \sigma^{a},\partial_{j}g_{0}\right\} +i\frac{\kappa_{sw}}{2}\epsilon_{kja}\left[\sigma^{a},g_{0}\partial_{j}g_{0}\right]\right)\label{eq:total_J}\; .
\end{equation}
Here $\theta$ is the  spin-Hall angle defined as  $\theta=2\lambda^{2}p_{F}/v_{f}\tau=l_{s0}/l$
and $\kappa_{sw}$ the ``swapping'' term $\kappa_{sw}=2\lambda^{2}p_{F}^{2}/3$\cite{lifshits2009swapping}.

Finally, by substituting Eq. (\ref{gani}) into Eq. (\ref{eq:kin_iso-2}) we obtain the Usadel equation:
\begin{equation}
\partial_{k}J_{k}+\left[(\omega-i{\bf h}\mathbf{\sigma)\tau_{3},}g_{0}\right]=-\frac{1}{8\tau_{SO}}\left[\sigma^{a}g_{0}\sigma^{a},g_{0}\right]+\epsilon_{kja}\frac{\lambda_{c}^{2}p_{F}}{4}v_{F}\left[\sigma^{a},\left\{ g_{0},\frac{1}{3}\partial_{k}g_{0}\partial_{j}g_{0}\right\} \right]\; .\label{eq:Usadel0}
\end{equation}
 Terms with two derivatives acting on the same $g$, i.e.
$\partial_{k}\partial_{j}g$ after summation over indices vanish because
of the antisymmetric tensor $\epsilon_{ijk}$. By substitution of Eq. (\ref{eq:total_J})
into Eq. (\ref{eq:Usadel0}) and going back to the real times representation
one obtains Eq. (5) of the main text. 

The generalisation of the Kupriyanov-Lukichev boundary condition at hybrid interfaces
 is straightforward from the current expression Eq.  (\ref{eq:total_J})  (we omit here the index $0$ in $G_0$):
\begin{equation}
\check{g}\partial_{k}\check{g}-\theta_{SH}\epsilon_{kja}\left\{ \sigma^{a},\partial_{j}\check{g}\right\} +i\kappa_{sw}\epsilon_{kja}\left[\sigma^{a},\check{g}\partial_{j}\check{g}\right]=-\frac{1}{2R_{b}\sigma_{F}}\left[g_{BCS},\hat{g}\right]\;,\label{eq:K-L}
\end{equation}

Observables like the charge current and magnetic moment can be expressed in terms of the quasiclassicla  Green's functions:
\begin{equation}
j_{k}=\frac{i\pi T}{16e}\sigma_{F}\sum_{\omega}{\rm Tr}\tau_{3}\check{J_{k}}\label{eq:current}
\end{equation}
and
\begin{equation}
m^{a}=\frac{\mu_{B}i\pi N_{0}T}{4}\sum_{\omega}Tr\tau_{3}\sigma^{a}\check{g}\;.\label{eq:magn_moment}
\end{equation}

{We should notice that in the normal case Eq. (\ref {eq:Usadel0}) simplifies drastically: First the retarded and advanced GFs equals to $\pm 1$ respectevely and hence 
there is only on equation for the Keldysh component which in such a case consist on the charge and spin distribution functions $g^K=f_c+f_s^a\sigma^a$. Second the equation can be straightfoirwarly integrated over energies and hence instead of writing the equations for $f_c$, $f_s^a$ one write them for the charge and the spin density $n\sim\int dEf_c(E)$ and $S^a\sim\int dE f_s^a(E)$. In particular by simple }

\subsection{Solution of the Usadel equation for a lateral multi-terminal S-F
structure}

Let us consider the geometry shown if Fig. 1 of the main text and
calculate the current through the $n$-th S/F interface, which is
given by Eq. (11) in the main text. Thus, we need to determine the
real part of the singlet component of the condensate induced in N.
In the geometry under consideration with an exchange field in $y$
direction, the anomalous GFs $\hat{f}=f_{s}+{\rm sgn}(\omega)\sigma^{y}f_{t}$
depends on two coordinates $x$ and $z$. It is convenient to introduce
the Fourier component $\hat{f}(q,z)$ with respect to $x$, 
\[
\hat{f}(x,z)=\int dqe^{iqx}\hat{f}(q,z)
\]
The singlet and triplet components then satisfy the following equations,
\begin{eqnarray}
\partial_{zz}^{2}f_{s}(q,z)-(k_{\omega}^{2}+q^{2})f_{s}(q,z)-ik_{h}^{2}f_{t}(q,z) & = & 0\label{eq:lin1}\\
\partial_{zz}^{2}f_{t}(q,z)-(k_{\omega}^{2}+q^{2}+k_{so}^{2})f_{t}(q,z)-ik_{h}^{2}f_{t}(q,z) & = & 0\label{eq:lin2}
\end{eqnarray}
 with boundary conditions at $z=0$
\begin{eqnarray}
\partial_{z}f_{s}(q,0)-iq\theta f_{t}(q,0) & = & if_{BCS}F_{0}(q)\label{eq:bc1}\\
\partial_{z}f_{t}(q,0)-iq\theta f_{s}(q,0) & = & 0\quad,\label{eq:bc2}
\end{eqnarray}
where $F_{0}(q)$ is the Fourier transform of the r.h.s of the boundary
condition at the S-electrodes described by 
\[
F_{0}(x)=\sum\gamma_{n}e^{i\varphi_{n}}\left[\Theta(x-x_{n})-\Theta(x-x_{n}-W_{n})\right]\quad,
\]
and $\gamma_{n}=1/R_{bn}\sigma_{0}$. Let us assume that $\varphi_{n}=0$
for all S terminals. According to Eq. (11) in the main text, to obtain
the current through the nth S/N boundary we only need to calculate
the real part of the singlet component, $f_{s}^{Re}(x)={\rm Re}f_{s}(x,0)$
at the S/F interface ($z=0$). One can straightforwardly verify from
Eqs. (\ref{eq:lin1})-(\ref{eq:bc2}) that in the linear order in
$\theta$ the Fourier component $f_{s}^{Re}(q)$ of $f_{s}^{Re}(x)$
is given by 

\[
f_{s}^{Re}(q)=ik_{h}^{2}\theta qF_{0}(q)s(q)\;,
\]
 where 

\begin{equation}
\frac{s(q)}{f_{BCS}}=\frac{1}{(k_{+}^{2}-k_{-}^{2})^{2}}\left[\frac{k_{-}^{2}-k_{\text{\ensuremath{\omega}}}^{2}}{2\left(q^{2}+k_{+}^{2}\right)}+\frac{k_{+}^{2}-k_{\text{\ensuremath{\omega}}}^{2}}{2\left(q^{2}+k_{-}^{2}\right)}-\frac{2k_{SO}^{2}}{\sqrt{q^{2}+k_{+}^{2}}\sqrt{q^{2}+k_{-}^{2}}}\right]\;,\label{eq:sq}
\end{equation}
 and $k_{\pm}^{2}=k_{\omega}^{2}+k_{so}^{2}/2\pm\sqrt{(k_{so}^{2}/2)^{2}-k_{h}^{4}}$.
Thus, $f_{s}^{Re}(x)$ can be obtained by transforming back 
\begin{equation}
f_{s}^{Re}(x)=k_{h}^{2}\theta\int dx_{1}\left[\partial_{x_{1}}F_{0}(x_{1})\right]s(x-x_{1})\label{eq:fsRe}
\end{equation}
 Since $F_{0}$ is a combination of step functions its spatial derivative
gives a sum of delta-functions, thus:
\begin{equation}
f_{s}^{Re}(x)=k_{h}^{2}\theta\sum_{n}\gamma_{n}\left[s(x-x_{n})-s(x-x_{n}-W_{n})\right]\;.\label{eq:lateral_fs1}
\end{equation}
 The inverse Fourier transform of the function $s(q)$, Eq. (\ref{eq:sq})
can be written explicitly as 
\begin{equation}
\frac{s(x)}{f_{BCS}}=\frac{k_{-}^{2}-k_{\omega}^{2}}{k_{+}(k_{+}^{2}-k_{-}^{2})^{2}}e^{-k_{+}|x|}+\frac{k_{+}^{2}-k_{\omega}^{2}}{k_{-}(k_{+}^{2}-k_{-}^{2})^{2}}e^{-k_{-}|x|}-\frac{2k_{so}^{2}}{\pi^{2}(k_{+}^{2}-k_{-}^{2})^{2}}\int dx_{1}K_{0}(k_{+}|x_{1}|)K_{0}(k_{-}|x-x_{1}|)\;,\label{eq:lateral_sx}
\end{equation}
 where $K_{0}$ is the modified Bessel function of second kind. Expressions
(\ref{eq:lateral_fs1}-\ref{eq:lateral_sx}) have been used to compute
the current from Eq. (11) in the main text. 

Now we consider a symmetric lateral structure with two S electrodes
(see Fig. 1c) of width $W$ at a distance $L$ from each other. We
assume that $h=0$ and a finite phase difference $\varphi$ between
the superconductors. According to Eqs. (\ref{eq:lin1}-\ref{eq:bc2})
the solutions for the singlet and triplet components are 
\begin{eqnarray}
f_{s}(q,z) & = & -\frac{if_{BCS}}{\kappa_{s}}F_{0}(q)e^{-k_{s}z}\label{eq:fs}\\
f_{t}(q,z) & = & -\frac{q\theta f_{BCS}}{k_{s}k_{t}}F_{0}(q)e^{-k_{t}z}\label{eq:ft}
\end{eqnarray}
where $k_{s}^{2}=k_{\omega}^{2}+q^{2}$and $k_{t}^{2}=k_{\omega}^{2}+k_{so}^{2}+q^{2}$
and $F_{0}(q)$ is the Fourier transform 
\[
F_{0}(x)=\gamma_{1}e^{-i\varphi/2}\left[\Theta(x+L/2+W)-\Theta(x+L/2)\right]+\gamma_{2}e^{i\varphi/2}\left[\Theta(x-L/2)-\Theta(x-L/2-W)\right]
\]
We calculate here the magnetic moment at $z=0$ that is given by 
\begin{equation}
m^{y}=2\pi\mu_{B}N_{0}T\sum_{\omega}{\rm Im}f_{s}^{*}(x,0)f_{t}(x,0)\label{eq:my}
\end{equation}
We need then to determine the Fourier transform of the prefactors
in Eq. (\ref{eq:fs}\ref{eq:ft}). In particular
\[
f_{s}(x,0)=-if_{BCS}\int dx'F_{0}(x')\frac{1}{\pi}K_{0}(k_{\omega}|x-x'|)
\]
 and 
\[
f_{t}(x,0)=i\theta f_{BCS}\int dx'\partial_{x'}F_{0}(x'){\cal F}(x-x')\;,
\]
with 
\[
{\cal F}(x-x')=\int\frac{dx''}{\pi^{2}}K_{0}(\sqrt{k_{\omega}^{2}+k_{so}^{2}}|x''|)K_{0}(k_{\omega}|x-x'-x''|)
\]
Substitution of these expressions into Eq.\ref{eq:my} gives 
\begin{eqnarray*}
m^{y}(x)=2\pi\gamma_{L}\gamma_{R}\mu_{B}N_{0}T\theta\sin\varphi\sum_{n} & \left\{ \int_{L/2}^{L/2+W}dx'K_{0}\left(k_{\omega}|x-x'|\right)\left[{\cal F}(x+\frac{L}{2}+W)-{\cal F}(x+\frac{L}{2})\right]\right. & -\\
 & -\left.\int dx'K_{0}\left(k_{\omega}|x-x'|\right)\left[{\cal F}(x-\frac{L}{2})-{\cal F}(x-\frac{L}{2}-W)\right]\right\} 
\end{eqnarray*}
This is the function plotted in the left panel of Fig. 2.

\end{widetext}

\end{document}